\documentclass[longauth,bibyear]{aa} % for the long lists of affiliations 
\usepackage[varg]{txfonts}

\usepackage{graphicx}
\usepackage{multirow}
\usepackage{rotfloat,amsmath}
\usepackage{color}

\begin{document} 

\newcommand{\abs}[1]{\ensuremath{\lvert#1\rvert}}
%Colors, spectrophotometry and albedo variations of 

   \title{
 Detection of exposed H$_2$O ice on the nucleus of comet 67P/Churyumov-Gerasimenko }
   \subtitle{as observed by Rosetta OSIRIS and VIRTIS instruments}

   \author{ M.A. Barucci  \inst{1} 
    \and G. Filacchione \inst{2}
        \and S. Fornasier \inst{1, 3}
             \and A. Raponi \inst{2}
              \and J.D.P. Deshapriya \inst{1}
               \and F. Tosi \inst{2}
                \and C. Feller \inst{1, 3}
     \and M.Ciarniello  \inst{2}
           \and H. Sierks \inst{4}
              \and F. Capaccioni \inst{2}  
              \and A. Pommerol \inst{5}
               \and M. Massironi\inst{6}  
           \and N. Oklay \inst{4}
            \and F. Merlin\inst{1,3}
               \and J.-B. Vincent \inst{4}
           \and M. Fulchignoni \inst{1,3}
                    \and A. Guilbert-Lepoutre \inst{7}
                 \and D. Perna \inst{1}    
              \and M.T. Capria \inst{2}  
              \and P. H. Hasselmann \inst{1}      
             \and B. Rousseau \inst{1}
              % \and OSIRIS and VIRTIS team to be added \inst{1,3}
                        \and C. Barbieri\inst{8}
                   \and D. Bockel\'ee-Morvan\inst{1}           
           \and P. L. Lamy\inst{9}
                  \and C. De Sanctis\inst{2}
           \and R. Rodrigo\inst{10, 11}
                  \and S. Erard\inst{2}
           \and D. Koschny\inst{12}
                  \and C. Leyrat\inst{1}
           \and H. Rickman\inst{13, 14}
                 \and P. Drossart\inst{1}
           \and H. U. Keller\inst{15}
               \and M. F. A'Hearn\inst{16}  
            \and G. Arnold\inst{17}
             \and J.-L. Bertaux\inst{18}
              \and I. Bertini\inst{19}
           \and P. Cerroni\inst{2}
     \and G. Cremonese\inst{19}
     \and V. Da Deppo\inst{20}
\and B. J. R. Davidsson\inst{21}
 \and M. R. El-Maarry \inst{5}
         \and S. Fonti \inst{22}
   \and M. Fulle\inst{23}
\and O. Groussin\inst{9}
\and C. G\"uttler\inst{4}
\and
S. F. Hviid\inst{17}
\and
W. Ip\inst{24}
\and L. Jorda\inst{9} 
 \and D. Kappel\inst{17}
\and J. Knollenberg\inst{17}
%\and
%G. Kovacs\inst{6}
\and J.-R. Kramm\inst{4}
\and E. K\"uhrt\inst{17}
\and M. K\"uppers\inst{25}
       %F. La Forgia\inst{8} 
\and  L. Lara\inst{26}  
\and M. Lazzarin\inst{19} 
\and  J. J. Lopez Moreno\inst{26}  
\and F. Mancarella\inst{22}
\and F. Marzari\inst{19}
%\and
%M. Massironi\inst{9,24}\and
%K.-D. Matz \inst{7} \and
%H. Michalik\inst{24}\and
%F. Moreno\inst{5}
\and S. Mottola\inst{17}
\and G. Naletto\inst{20}
\and M. Pajola\inst{27} 
      \and E. Palomba\inst{2}
\and E. Quirico\inst{28} 
\and B. Schmitt\inst{28} 
\and N. Thomas\inst{5}
\and C. Tubiana\inst{4} 
% 
%J.-B. Vincent\inst{6}
 }

   \institute{LESIA, Observatoire de Paris, CNRS, UPMC Univ. Paris 06, Univ. Paris-Diderot, 5 Place J. Janssen,  92195 Meudon Principal Cedex, France 
              \email{antonella.barucci@obspm.fr}
                  \and INAF-IAPS, 00133 Rome, Italy
                \and Univ Paris Diderot, Sorbonne Paris Cit\'{e}, 4 rue Elsa Morante, 75205 Paris Cedex 13, France    
                  \and Max-Planck-Institut f\"ur Sonnensystemforschung, Justus-von-Liebig-Weg, 3 37077 G\"ottingen, Germany 
                    \and Physikalisches Institut, Sidlerstrasse 5, University of Bern, CH-3012 Bern, Switzerland
                   \and    Dipartimento di Geoscienze, University of Padova, Italy %Massironi
                   \and Observatoire des Sciences de l'Univers, Besan\c{c}on, France
                 \and Department of Physics and Astronomy "G. Galilei", University of Padova, Vic. Osservatorio 3, 35122 Padova, Italy
                   \and Laboratoire d’Astrophysique de Marseille UMR 7326, CNRS \& Aix Marseille Universit\'e, 13388 Marseille Cedex 13, France
             \and  Centro de Astrobiolog\'ia, CSIC-INTA, 28850 Torrej\'on de Ardoz, Madrid, Spain %Rodrigo
                \and International Space Science Institute, Hallerstrasse 6, 3012 Bern, Switzerland %Rodrigo
                   \and Research and Scientific Support Department, European Space Agency, 2201 Noordwijk, The Netherlands % Koschny
                \and Department of Physics and Astronomy, Uppsala University, 75120 Uppsala, Sweden % Davidsson, Rickman
                \and
              PAS Space Reserch Center, Bartycka 18A, 00716 Warszawa, Poland  %Rickman    
              \and
                Institute for Geophysics and Extraterrestrial Physics, TU Braunschweig, 38106 Braunschweig, Germany % Keller     
              \and
                Department for Astronomy, University of Maryland, College Park, MD 20742-2421, USA %A'He  
               \and Institute of Planetary Research, DLR, Rutherfordstrasse 2, 12489 Berlin, Germany     
                \and
                LATMOS, CNRS/UVSQ/IPSL, 11 Boulevard d'Alembert, 78280 Guyancourt, France % Bertaux   
                  \and INAF--Osservatorio Astronomico di Padova, Vicolo dell'Osservatorio 5, 35122 Padova, Italy % Bertini
                      \and
                Department of Information Engineering - University of Padova, Via Gradenigo 6, 35131 Padova, Italy % Naletto
                      \and
                JPL, 4800 Oak Grove Drive, Pasadena, CA91109, USA
                       \and Dipartimento di Fisica, Universita del Salento, Italy % Fonti
                   \and
                INAF -- Osservatorio Astronomico di Trieste, via Tiepolo 11, 34143 Trieste, Italy %Fulle
                \and
                Institute for Space Science, National Central University, 32054 Chung-Li, Taiwan
           \and
                ESA/ESAC, PO Box 78, 28691 Villanueva de la Ca\~nada, Spain %Kueppers      
                 \and Instituto de Astrof\'isica de Andaluc\'ia -- CSIC, 18080 Granada, Spain 
               \and
               NASA Ames Research Center, CA 94035, USA % pajola
                   \and
               UJF-Grenoble 1/CNRS-INSU, France
                %INAF--Osservatorio Astronomico di Padova, Vicolo dell'Osservatorio 5, 35122 Padova, Italy % Cremonese
                %\and
                %CNR--IFN UOS Padova LUXOR, Via Trasea 7, 35131 Padova, Italy % Da Deppo
                %\and
                %Department of Mechanical Engineering -- University of Padova, Via Venezia 1, 35131 Padova, Italy % Debei
                %\and
                %UNITN, Universit\'a di Trento, Via Mesiano, 77, 38100 Trento, Italy % De Cecco
                %\and
                %Institut f\"ur Datentechnik und Kommunikationsnetze, 38106 Braunschweig, Germany % Michalik
                %\and
                %Department of Information Engineering - University of Padova, Via Gradenigo 6, 35131 Padova, Italy % Naletto
                %\and Center of Studies and Activities for Space (CISAS) "G. Colombo", University of Padova, Via Venezia 15, 35131 Padova, Italy
                 %\and
                %Physikalisches Institut, Sidlerstrasse 5, University of Bern, CH-3012 Bern, Switzerland %Thomas
                %\and
                % Planetary and Space Sciences, Department of Physical Sciences, The Open University, Walton Hall, Milton Keynes, MK7 6AA, UK %colin   
                 }

   \date{Received 21 April 2016}

\newpage

% \abstract{}{}{}{}{} 
% 5 {} token are mandatory
 
  \abstract
  % context heading (optional)
  % {} leave it empty if necessary
    {Since the orbital insertion of the Rosetta spacecraft,  comet 67P/Churyumov-Gerasimenko
(67P/C-G) has been mapped by OSIRIS camera and VIRTIS spectro-imager,  producing a huge quantity of images and spectra of the comet's nucleus.}
  % aims heading (mandatory)
   {The aim of this work is to search for the presence of H$_2$O on the nucleus which, in general, appears very dark and rich in dehydrated organic material.   After  selecting  images of the bright spots which could be good candidates to search for H$_2$O ice, taken at high resolution by OSIRIS, we check for  spectral cubes of the selected coordinates to identify these spots observed by VIRTIS.}
  % methods heading (mandatory)
   { The selected OSIRIS images were processed with the OSIRIS standard pipeline and corrected for the illumination conditions for each pixel using the Lommel-Seeliger disk law. The spots with higher I/F were selected and then analysed  spectrophotometrically and compared with the surrounding area. We selected 13 spots as good targets to be analysed by VIRTIS to search for the 2 $\mu$m absorption band of water ice in the  VIRTIS spectral cubes. } 
  % results heading (mandatory)
   {Out of the 13  selected bright spots,  eight of them present positive H$_2$O ice detection on the VIRTIS data.  A spectral analysis was  performed and the approximate temperature of each spot was computed.  The H$_2$O ice content  was confirmed  by modeling the spectra with mixing (areal and intimate) of  H$_2$O ice  and  dark terrain, using Hapke's radiative transfer modeling. We also present  a detailed analysis of the detected spots.  } 
  % conclusions heading (optional), leave it empty if necessary 
   {}

\titlerunning{ H$_2$O ice spots of the 67P comet nucleus}
%\authorrunning{}
   \maketitle
%
%________________________________________________________________

\section{Introduction}

Space exploration has triggered major progress in our understanding of comets beginning in March 1986 with the exploration of comet 1P/Halley by an armada of missions including the ESA Giotto mission (Reinhard and Battrick, 1986). With the arrival of the ESA Rosetta mission at the comet 67P/Churyumov-Gerasimenko (67P/C-G) on July 2014, comets appear more complex and fascinating than ever.  All the visited comets  show a low visible albedo and heterogeneous surface (Barucci et al. 2011). However 67P/C-G and some other periodic comets  reveal the presence of intriguing bright spots on the surface (Sunshine et al. 2006; Sunshine et al. 2012; and Li et al. 2013; Pommerol et al. 2015). To better understand  the properties and composition of the comet 67P/C-G is one of the major objectives of the ESA Rosetta mission, and all on-board instruments have so far contributed  with  high quality and a precious quantity of data. 

Since the orbital insertion of the Rosetta spacecraft, the comet nucleus has been mapped by both OSIRIS (Optical, Spectroscopic, and Infrared Remote Imaging System), and VIRTIS (Visible InfraRed Thermal Imaging Spectrometer) acquiring a huge quantity of surface images in different wavelength bands and spectra, and producing the most detailed maps at the highest spatial resolution of a cometary nucleus surface.  The OSIRIS imaging system (Keller et al. 2007)   is composed of the Narrow Angle Camera (NAC) designed to study the nucleus with 11 large band filters at different wavelengths from the ultraviolet (269 nm) to the near-infrared (989 nm), while the Wide Angle Camera (WAC) is devoted to the study of gaseous species in the coma with a set of 14 narrow band filters ranging from the ultraviolet to visible wavelengths.  The OSIRIS imaging system was the first instrument capable of mapping a comet surface at high resolution, reaching a  maximum resolution of 11cm/px during the closest fly-by that occurred on February 14, 2015, at a distance of $\sim$  6 km from the nucleus surface.  VIRTIS (Coradini et al. 2007) is composed of two channels: VIRTIS-M, a spectro-imager operating both in the visible (0.25$-$1.0 $\mu$m) and infrared (1.0$-$5.0 $\mu$m) ranges at low spectral resolution ($\lambda/\delta\lambda$=70--380), devoted to surface composition, and VIRTIS-H, a single-aperture infrared spectrometer (1.9$-$5.0 $\mu$m) with higher spectral resolution capabilities ($\lambda/\delta\lambda$=1300-3000) devoted to  the investigation of activity.

     \begin{figure*}[t]
     \includegraphics[width=1\textwidth]{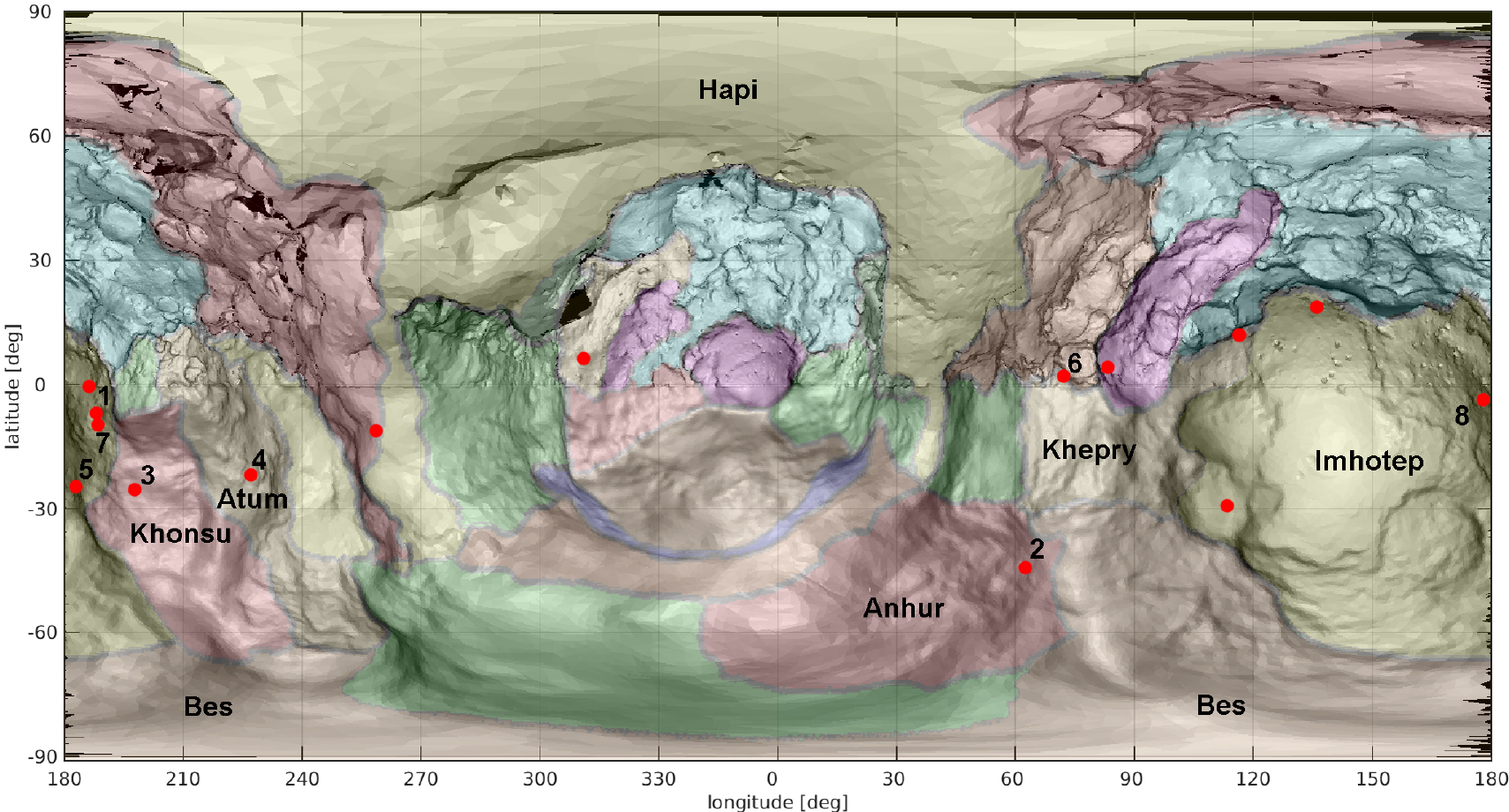}
      \caption{Map of comet 67P/Churyumov-Gerasimenko, resulting from merging a more detailed shape model  SHAP4S (Preusker et al. 2015) for the northern hemisphere and shape model SHAP5 (Jorda et al. 2016) for the southern hemisphere. In red the selected bright spots are reported, based on OSIRIS images  and a spectro-photometric analysis, considered as good targets to be investigated by  an analysis of VIRTIS data, plus the two bright spots analysed by Filacchione et al. (2016a). The numbers (1-8) represent the spots with positive detection of H$_2$O ice by VIRTIS analysis discussed in this paper. }
         \label{spots}
   \end{figure*}

%%%%%%%%%%%%%% 

The OSIRIS images of 67P/C-G show a highly shaped, irregular bilobed comet, with a dark, dehydrated, and morphologically complex surface characterized by several terrain types, including numerous  diverse geomorphologic features (Sierks et al. 2015). The comet's surface is highly heterogeneous with different geological terrains showing smooth, dust-covered areas, large scale depressions, brittle materials with many pits and circular structures, and exposed consolidated areas (Thomas et al. 2015a; El-Maarry et al. 2015; El-Maarry et al. 2016). Pits have also been connected to activity, possibly accompanied by outbursts (Vincent et al. 2015). The complex surface of comet 67P/C-G shows regions  covered by different layers of dust on  both lobes, including areas with evidence of transport and redistribution of dust materials (Thomas et al. 2015b).  Temporal variations of morphological structures have also been  observed on the smooth terrains of the Imhotep region  (Groussin et al. 2015), as well as in other regions (Fornasier et al. 2016), in particular when the comet was close to perihelion. The comet shows albedo variation of up to about 25\% and spectrophotometric analysis  (Fornasier et al. 2015)  identified at least three groups of terrains with different spectral slopes (computed in the 535-882 nm range). These differences have been associated with the local composition variation, but since many different surface characteristics overlap, this makes the interpretation difficult.  Oklay et al. (2016a) also studied  surface variegation on the comet, detecting local color inhomogeneities
connected to active and inactive surface regions.

\begin{table*}
\begin{center}
\caption{Observing conditions for the OSIRIS images as reported in Fig. 2, where the ice spots have been identified.  The time (UT) refers to the start time of the first image of each sequence, followed by the number of filters available. The diameter size (d) of the spots along with the location region, phase angle ($\alpha$), 
 distance between Rosetta spacecraft, and comet surface ($\Delta$), spatial resolution (R), latitude (Lat), and longitude (Long) are reported.}
\small{
\label{observing}
\begin{tabular}{l l l c c c c c c c} 
 \hline
N. & Time reference &   Filters &   d    & Region &  $\alpha$ & $\Delta$ & R & Lat  &  Long   \\ 
 &                              &               & (m)    &   &($^{\circ}$)            &      (Km)         &      (m/px)   &    ($^{\circ}$)   & ($^{\circ}$)  \\   \hline
1 & 2015-06-27T13h26 & F22, F23, F41, F24, F71, F27, F51, & 36   &  Imhotep  & 89.50 &  191.94 & 3.6 &-5.8 & 189.4 \\
& & F61, F28, F15& & & & & & &   \\
2   &   2015-06-27T17h48    &  F22, F23, F41, F24, F71, F27, F51,    &45      &  Anhur  &  89.39 & 188.43  & 3.5 & -41.7 &      63.7     \\
&&F61, F28, F15  & & & & & & &   \\
 3 &        2015-04-12T21h42      & F22, F23, F41, F24, F71, F27, F51  &11   &  Khonsu      &     80.46     &    147.98      & 2.7 &  -23.8 & 198.3               \\  
 &&F61, F28,F16,  F15  & & & & & & &   \\
4 &  2014-11-22T04h57& F22, F23, F24,  F27, F28, F51, F61   &10  &       Atum               & 92.70    &  29.50 & 0.54 & -20.7 & 227.4   \\
5 &   2014-11-22T06h32& F22, F23, F24,  F27, F28, F51, F61  &6.5 &      Imhotep                   &  92.78  &  29.50  & 0.54 &  -22.0 & 182.8   \\  
6 &  2014-09-19T09h19 & F22, F16, F23,  F24, F41   & 2-5 (each) &        Khepry            &    70.48 & 26.50  &  0.49 &  4.2 &  71.7  \\  
7 &  2014-09-05T05h21 &  F22, F23, F27, F16, F28, F41, F71  & 3-5 (each) &     Imhotep                     &  57.23  &  41.44 & 0.77 & -8.1 &188.3  \\ 
8 & 2014-09-05T08h00&   F22, F23, F27, F16, F28, F41, F71&  6 & Imhotep & 58.43 & 40.76 & 0.75 & -2.4 &174.8\\ 
\hline \hline
\end{tabular}
}
\end{center}
\end{table*}
The first results by VIRTIS (Capaccioni et al. 2015) about the spectral analysis showed the presence of a broad absorption feature around  2.9--3.6 
$\mu$m present across the entire observed region and compatible with carbon-bearing compounds (opaque minerals associated with organic macromolecular materials) with no evidence of ice-rich patches.  Later on, De Sanctis et al. (2015) detected the first evidence for the presence of H$_2$O ice as part of a diurnal cycle on the neck of the comet, while Filacchione et al. (2016a) identified H$_2$O ice on two gravitational debris falls in the Imhotep region exposed on the walls of elevated structures.  The latter was interpreted as being possibly  extended layering in which the outer dehydrated crust is superimposed over water ice-enriched layers. During the first mapping phase of 67P/C-G nucleus, completed in August-November 2014 (heliocentric distances between 3.6 and 2.7 AU), VIRTIS-M  achieved a complete mapping of the illuminated regions in the equatorial and northern hemisphere, which  enabled us to retrieve the first compositional maps by using VIS and IR spectral parameters (Filacchione et al. 2016b). During the same period, coma observations performed by VIRTIS-M (Migliorini et al. 2016) and VIRTIS-H (Bockel\'ee-Morvan et al. 2015) channels have traced the H$_2$O vapor emission, which occurs preferentially above the illuminated regions of the northern hemisphere. 

As limited evidence of exposed H$_2$O ice regions  has so
far been collected, the aim of this work is to investigate in depth the composition of the 67P/ C-G surface, combining the high spatial resolution images of OSIRIS and the high spectral resolution of VIRTIS for detecting and emphasizing interesting ice spectral signatures.  Over 100 meter-sized spots  were  identified by Pommerol et al (2015),  possibly associated with the presence of H$_2$O, on the basis of laboratory experiments, but with no  confirmation of the real presence of ice. Deshapriya et al. (in preparation) are collecting  a catalogue of large bright spots that are present on the surface of the comet by analyzing the OSIRIS images and spectrophotometry data.
For this work we selected the largest spots, good candidates to search for H$_2$O ice that  could be detected at the lower angular resolution of VIRTIS. We identify large features with high albedo and low spectro-photometric slope with OSIRIS, we compute accurate coordinates, and we analyze them on the basis of the VIRTIS spectra. Over the large number of spots identified by OSIRIS, 13 of them were checked by VIRTIS. Eight  of them show clear evidence of H$_2$O ice in their spectra. 

In this paper we report on the analysis of the eight bright spots for which we obtained a positive detection of H$_2$O ice in VIRTIS data. In Section 2, the OSIRIS data and the performed analysis are presented and, in Section 3, the VIRTIS data, while in Section 4 the spectral modeling of the selected spots is described. In Section 5,   a detailed analysis of the spots and surrounding area is reported, while in Section 6, a possible evolution of the area is discussed. The main aim of this work is to confirm the unambiguous presence of H$_2$O ice by spectral analysis.

   \begin{figure*}
   \centering
\includegraphics[width=1\textwidth]{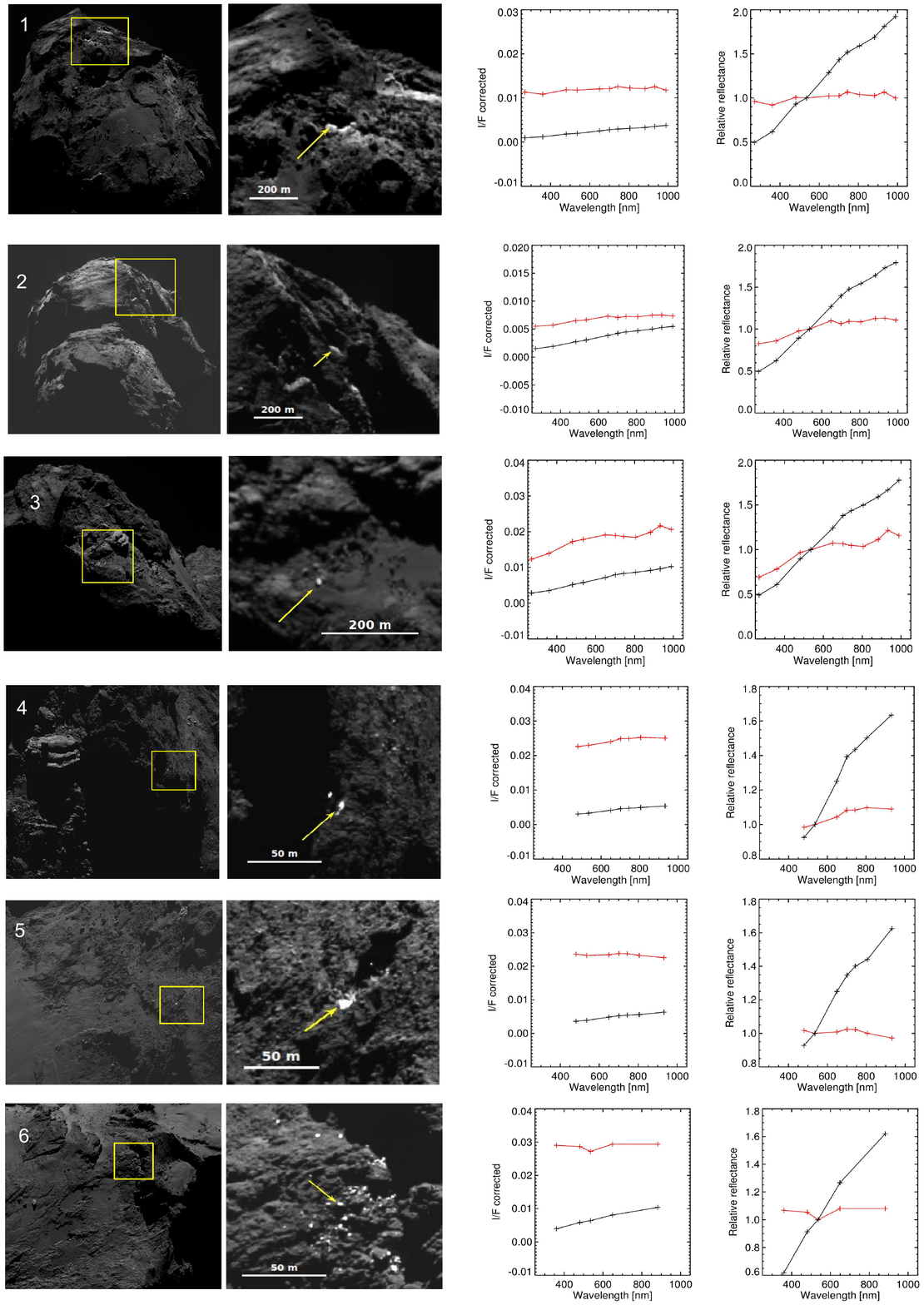}
\clearpage
 \end{figure*}
\begin{figure*}
\centering
\includegraphics[width=1\textwidth]{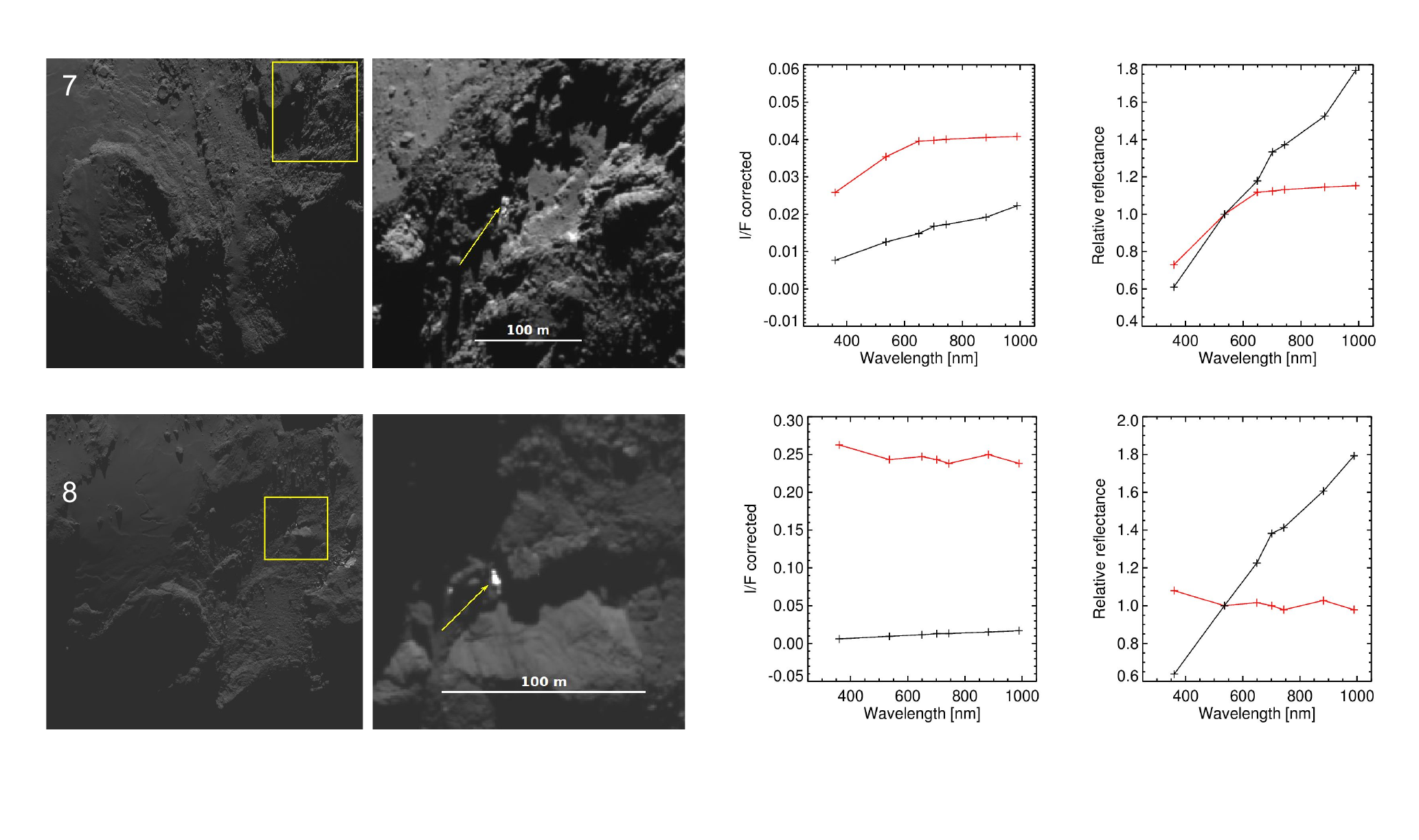}
%                       \includegraphics[width=0.3\textwidth]{fig2_bis.pdf}
%\caption{NAC OSIRIS images for the spot N.7 and N.8 reported in table 1 with the zoom of the bright spots (on the right). }
\caption{NAC OSIRIS images (first column) for the eight spots reported in Table 1 with a zoom on the spot (second column). The images have been taken with F22 filter (at 649.2 nm). The arrows indicates the spots that have been analysed using boxes of 3x3 pixels. The measured  $I/F(\alpha)$  of the bright spots are reported in red, while the surrounding area is reported in black (third column). The  relative reflectance (normalized to F23 at 535 nm) of the indicated bright spot in red and the surrounding area in black are represented in the fourth column. }
         \label{images}
   \end{figure*}

  \section{OSIRIS data}
 Bright spots have been observed on 67P/C-G at various locations on the nucleus throughout the cruise phase of the Rosetta mission. First detections of these spots date back to as early as August 2014 and they continued to appear in numerous forms, be it an ensemble of a plethora of small bright spots, or much larger individual or twin bright patches. The abundance of these spots reached a peak during the perihelion passage of comet 67/P C-G in August 2015, resulting in the largest white patches ever detected on the cometary nucleus.  The first objective of  this study is to explore the nature of these spots in terms of spectrophotometry. 

        We started by selecting potential bright spots found on OSIRIS NAC data  with good spectro-photometry and spatial coverage.  In the event of an observational sequence of several filters, owing to the fact that both the spacecraft and cometary nucleus are constantly in motion, the recorded images do not necessarily show exactly the same field of view. Owing to the time difference between a pair of consecutive images being around ten seconds, there is a small shift in the fields of view. Hence we have to take this shift into account when stacking up images and creating a data cube for spectral analysis. To achieve this, we adopted an algorithm that  automatically identifies identical features in consecutive images and estimates the affine transformation matrix between each couple of consecutive images (ORB and RANSAC tools implemented by van der Walt et al., 2014).  Then we analyzed the data cubes to identify bright spots. The  basic criterion to identify these spots was their high reflectance compared to the typical nucleus in the filter wavelengths observed. We note that the  data used are from 3B level of the standard OSIRIS data reduction pipeline, which accounts for the correction for bias, flat field, geometric distortion, solar flux, and calibration in absolute flux (Tubiana et al. 2015).  Thus the data available in the reduced files are in the form of radiance factor that corresponds to the ratio between the observed scattered radiance (I) from the comet and incoming solar irradiance ($F_{\lambda}$) at the heliocentric distance of the comet, being referred to as the $I/F$ value 
  
\begin{equation}
I/F(\lambda)  = \frac{\pi I(i,e,\alpha,\lambda)}{F_{\lambda}},
\end{equation}

where  $i$ is the incident angle, $e$ is the emission angle, and $\alpha$ is the phase angle.

Next we proceeded to generate synthetic images to correct for the illumination conditions and observational geometry using the OASIS tool  (Jorda et al. 2010) and the shape model SHP5 (Jorda et al. 2016).  This yields incident and emission angles for each pixel, which enables us to apply the Lommel-Seeliger disk law

   \begin{equation}
D(i,e) =  \frac{2 \cos(i)}{\cos(i) + \cos(e)}
\end{equation} 

\begin{equation}
(I/F)_{corr}(\alpha, \lambda) = \frac{\pi I(i,e,\alpha,\lambda)/(F_{\lambda})}{D(i,e)}
.\end{equation}

Derived from I/F plots we also produced relative reflectance plots with radiance factors normalized to a given filter 
(i.e. F23, green filter at 535 nm), enabling us to get an insight into the composition of the areas sampled for this analysis. As the ice displays a flatter spectrum in comparison to the red nature of organic-rich typical comet's nucleus material,  we set the following criteria for the bright spots to qualify as final candidates
: i) higher albedo properties than the typical nucleus  I/F and ii) flat spectral behavior, compared to the typical nucleus on relative reflectance plots. 

The above method enabled us to filter  potential candidates and discard certain previously catalogued bright spots that had shown higher albedo properties, but not necessarily having flat spectral behavior when normalized, thus failing to meet the second requirement. In this case, the high albedo properties were probably  due to the illumination conditions during the observation. 
 
We selected 13 spots (or cluster of spots) reported in the map (Fig. 1)  as the best sample to be analysed by spectroscopy with VIRTIS. In this paper we present only  eight spots (Table 1) for which VIRTIS spectral analysis  gave positive detection of H$_2$O ice signatures. The OSIRIS images of the area and the zoom of the selected eight bright spots were reported  in Fig. 2, together with the measured I/F and the relative spectro-photometry reflectance.  As shown in  Fig. 2, the spots 4 (depending on the shadow), 6, and 7 belong to a cluster.

\begin{table*}
        \centering
        \caption{Coverage of OSIRIS bright spot locations in VIRTIS dataset and [number] of 2 $\mu m$ absorption band positive identifications. MTPs are Rosetta mission medium term plan time intervals, each one having a duration of about one month (MTP7 corresponds to September 2014, MTP15 to April 2015). In the last column, the total positive detections are reported.}
\begin{tiny}
        \begin{tabular}{lccccccccccc}
                \hline
OSIRIS Spot &Ice spot N. & MTP007 & MTP008 & MTP009 & MTP010 & MTP011 & MTP012 & MTP013 & MTP014 & MTP015 & Total\\
                \hline
                
nl1-5b&1 & 20 [0] & 0 & 0 & 0 & 0 & 0 & 15 [2] & 139 [0] & 89 [0] & \textbf{2} \\
nl2-7b&-&  0 & 0 & 0 & 40 [0] & 0 & 16 [0] & 15 [0] & 8 [0] & 8 [0] & \textbf{0}\\
nl3-12b&-& 149 [0] & 0 & 36 [0] & 144 [0] & 64 [0] & 0 & 32 [0] & 77 [0] & 87 [0] & \textbf{0} \\
nl4-13&2 & 0 & 0 & 0 & 0 & 0 & 0 & 4 [0] & 106 [9] & 49 [0] & \textbf{9}\\
nl5-16&-& 142 [0] & 0 & 0 & 32 [0] & 26 [0] & 0 & 45 [0] & 121 [0] & 70 [0] & \textbf{0} \\
nl6-17a&3 & 0 & 0 & 0 & 0 & 0 & 0 & 15 [0] & 116 [9] & 64 [3] & \textbf{12}\\
nl7-19 & 4& 0 & 0 & 195 [32] & 0 & 0 & 0 & 29 [0] & 52 [0] & 39 [0] & \textbf{32}\\
nl8-20 & 5&40 [0] & 0 & 0 & 300 [86] & 0 & 0 & 38 [0] & 190 [0] & 87 [0] & \textbf{86}\\
nl9-22 & - &225 [0] & 0 & 140 [0] & 86 [0] & 0 & 0 & 21 [0] & 77 [0] & 59 [0] & \textbf{0}\\
nl10-23b &-& 0 & 0 & 0 & 0 & 0 & 27 [0] & 184 [0] & 58 [0] & 27 [0] & \textbf{0}\\
nl11-14b & 6&102 [3] & 711 [0] & 72 [0] & 140 [0] & 70 [0] & 0 & 46 [0] & 76 [0] & 65 [0] & \textbf{3}\\
nl12-24 & 7&94 [10] & 0 & 0 & 0 & 0 & 0 & 14 [0] & 122 [4] & 66 [1] & \textbf{15}\\
nl13-25 & 8&326 [0] & 0 & 0 & 345 [3] & 41 [0] & 0 & 43 [0] & 93 [0] & 85 [0] & \textbf{3}\\
\hline
\end{tabular}
\end{tiny}
\label{tbl:table_virtis_listing}
\end{table*}

%
  %          \begin{figure*}[t]
 %\includegraphics[height=1\textwidth,  angle=-90]{15spots.pdf}
    %  \caption{Complete map of comet 67P/Churyumov-Gerasimenko which is a results of merging a more precise shape model  SHAP4S for the hemisphere Nord and shape model SHAP5 for the hemisphere Sud}
       %  \label{spots}
  % \end{figure*} 

  \begin{table*}
%\begin{sideways} 
\begin{tiny}
\begin{center}
\caption{ Summary of VIRTIS-M dataset processed in this work. For each spot, we report  the observations offering the best signal-to-noise conditions with the pixel position (sample and line) reported in Table 4. For each pixel, basic information about observation time, geometry conditions, distance between the Rosetta spacecraft and comet surface ($\Delta$), Local Solar Time (LST), and retrieved temperature (T) are given. The integration time is 3s for all data reported. The cube parameters indicate the size of the acquisition in bands, sample, line dimensions.}

\begin{tabular}{lllcccccccc}
                
                \hline
N. & Observation &  Cube & Start Time & End Time  & Phase  & Incidence  &  Emission &  $\Delta$  & LST & T  \\ 
& name & Parameters & (UT)   &  &  (deg) &  (deg) &  (deg) &  (km) & (hr) & (K)  \\ 
                \hline
1 & I1$\_$00383518966 & 432, 256, 158 & 2015-02-25T21:04:00 & 2015-02-25T21:30:19  & 53.02 & 66.94 & 41.92 & 81.51 & 12.23 & 203 \\ 
2 & I1$\_$00385906923  & 432, 256, 70  & 2015-03-25T12:23:18 & 2015-03-25T12:46:34 & 73.49 & 67.08 & 45.83 & 88.23 & 15.20 & 197 \\
3 & I1$\_$00385885107  & 432, 256, 70  & 2015-03-25T06:19:42 & 2015-03-25T06:42:58  & 74.47 & 53.03 & 31.27 & 94.06 & 12.43 & 218  \\
4 & I1$\_$00373462192  & 432, 256, 86  & 2014-11-01T11:31:03 & 2014-11-01T11:45:22  & 103.07 & 60.11 & 43.28 & 32.39 & 11.04 & 168 \\
5 & I1$\_$00377182711  & 432, 256, 80  & 2014-12-14T12:59:43 & 2014-12-14T13:13:02  & 91.77 & 43.41 & 52.15 & 19.44 & 15.44 & 188 \\ 
6 & I1$\_$00376302211  & 432, 256, 80  & 2014-12-04T08:24:43 & 2014-12-04T08:38:01  & 91.07 & 57.34 & 70.64 & 23.49 & 14.81 & 179 \\
7 & I1$\_$00369356914  & 432, 256, 109 & 2014-09-14T23:09:43 & 2014-09-14T23:45:57  & 66.89 & 78.22 & 30.26 & 28.16 & 11.03 & 163 \\
8 & I1$\_$00377184571  & 432, 256, 74  & 2014-12-14T13:30:43 & 2014-12-14T13:43:02  & 92.66 & 48.30 & 54.87 & 19.92 & 15.59 & 158 \\
\hline 
\hline
\label{table:virtis_obs}
\end{tabular}
\end{center}
\end{tiny}
%\end{sideways}
\end{table*}

 \begin{figure}
 \centering
\includegraphics[width=9.0cm,angle=0]{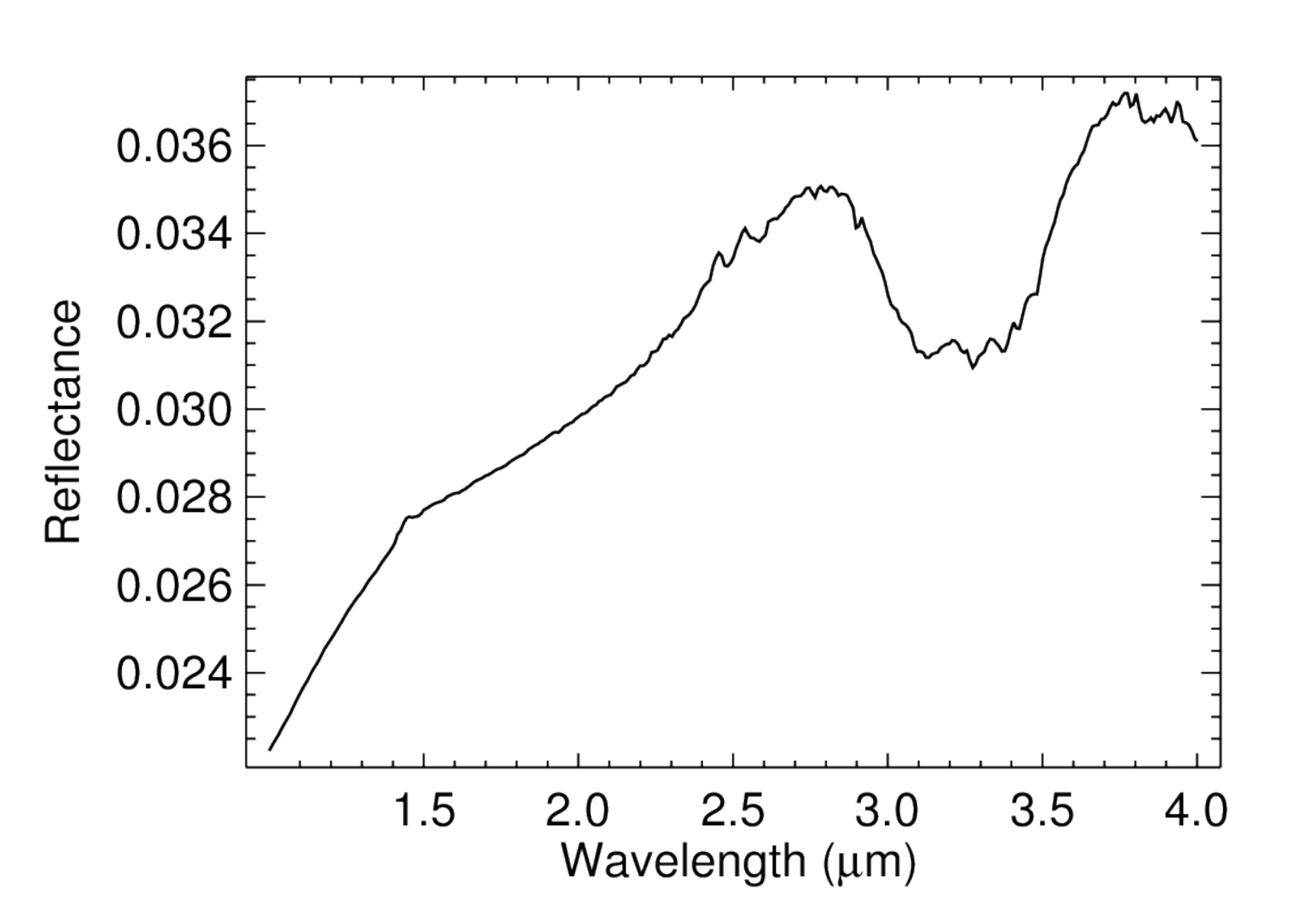}
\caption{Spectrum of Dark Terrain unit that corresponds to the average spectrum of the comet's surface. }
 \label{xx}
\end{figure}

\section{VIRTIS data}
The search of VIRTIS-M spectra in correspondence with the bright albedo features identified on OSIRIS images  needed a deep data mining of the dataset. Since the entire nucleus was imaged with  very high redundancy from a wide range of distances, local times and illumination/viewing geometries, the research has been performed starting from georeferenced data. For each individual pixel in each VIRTIS-M observation, many geometry parameters, including longitude, latitude, incidence, emission, phase angles, distance, and local solar time for the pixel center and four corners, were computed by means of SPICE (Acton, 1996) routines that are able to reconstruct these quantities starting from spacecraft and comet attitude and trajectory kernels. Once computed and validated by the VIRTIS team, these geometry (.GEO) cubes are released through ESA's PSA archive and made publicly available. 
The nucleus shape model used for computation is SHAP5, derived from OSIRIS images by using a stereophotoclinometry method (Jorda et al. 2016). The coordinates grid is based on the Cheops frame (Preusker et al. 2015). Some examples of geometry parameters computed for VIRTIS-M nucleus observations are given in Filacchione et al. (2016b).
The geometry information relative to each pixel is therefore ingested in a database.

\begin{figure*}
   \centering
 \includegraphics[width=0.8\textwidth]{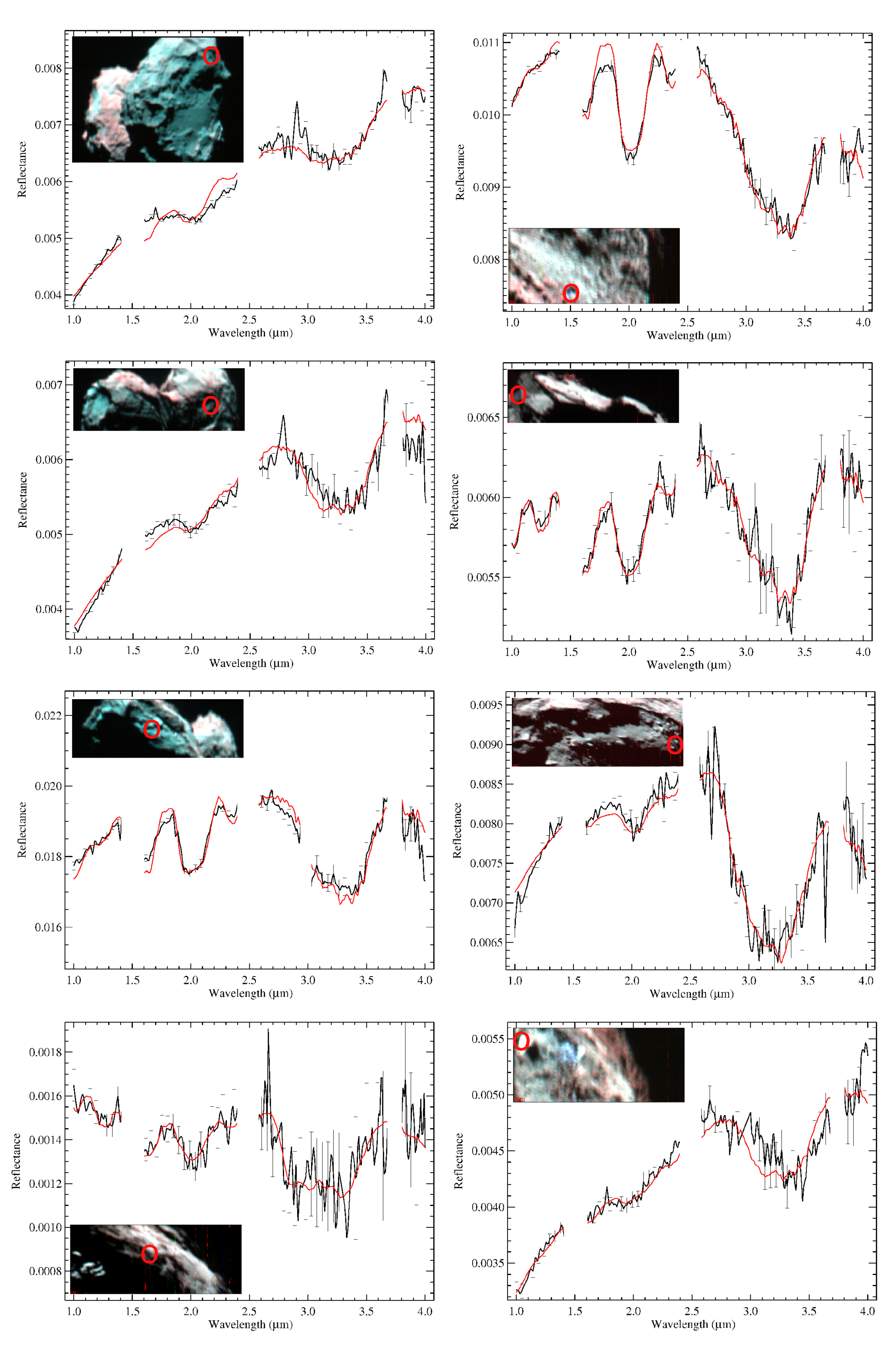}
      \caption{VIRTIS-M infrared data of the eight spots. Spots 1-4 are shown in the left column from top to bottom. Spots 5-8 in the right column from top to bottom. Infrared images in the insert of the plot are built from a combination of spectral bands taken at 1.32 $\mu m$ (B channel), 2.0 $\mu m$ (G), 4.0 $\mu m$ (R). For each spot, we report the VIRTIS-M observed reflectance (black curve), not corrected for phase angle, and best fit (red curve) as derived from the pixels reported in Table \ref{table:virtis_obs} and indicated by  red circles on the images. The gaps in the spectral ranges correspond to order-sorting filter-junction wavelengths that can produce unreliable features and, for this reason, are not taken into account in the analysis.}
         \label{Spot}
   \end{figure*}

\subsection{Search for the 2 $\mu m$ absorption feature}
Using the method  described above, the search of VIRTIS-M pixels located in correspondence with the bright spot coordinates identified on OSIRIS images was performed. As a general rule, we  selected all VIRTIS-M pixels within a radius of $2^\circ$ in longitude and latitude around the position estimated on the OSIRIS images. The corresponding reflectance spectra are grouped together to form a spectrogram (Filacchione et al., 2016a), one for each MTP, and then further processed by calculating the 2 $\mu m$ band depth, used as a proxy to determine the presence of H$_2$O ice on the surface in the case of values that were larger than a  5$\%$ threshold.  Water ice reflectance shows diagnostic absorptions at 1.5, 2.0 and 3.0 $\mu m$. The  decision to use only the 2 $\mu m$ band as a proxy to identify the presence of water ice on 67P/C-G surface was driven by two different requirements: i) the 1.5 $\mu m$  band is partially corrupted by the presence of an instrumental order-sorting filter, which makes it difficult to retrieve a correct band shape, particularly for pixels close to sharp illumination transitions and shadows; ii) the intense 3 $\mu m$ band has a complex shape owing to the overlapping of the water and organic material absorptions, which causes changes in shape, center, and depth, depending on the relative abundances of the two end-members. Conversely, the 2 $\mu m$ spectral range is not influenced by similar effects. Moreover, the 2 $\mu m$  band is well-defined for a wide range of grain sizes, making it a good spectral marker to identify the presence of water ice.

As a general rule, the detection of the H$_2$O ice on a given place can be limited by unfavorable instrumental signal-to-noise conditions, spatial resolution on ground and oblique illumination/viewing geometry. A summary of the observations showing a positive identification of the 2 $\mu m$ H$_2$O ice-band feature is given in Table 2. 
Infrared color images of the eight spots are shown in Fig. 4, together with the observed reflectance spectra and best spectral fits, as discussed later in Section 4. In absence of water ice, VIRTIS  spectra correspond with the Dark Terrain unit, as reported in Fig. 3, which shows a featureless red slope  in the 1-2.6 $\mu m$ range and an intense organic material absorption band at 3.0 $\mu m$.

\subsection{Temperature of the icy spots}
%For each spot we have identified VIRTIS-M observations showing the eight spot areas imaged with the best illumination and viewing geometries to maximize the signal to noise ratio. The VIRTIS-M dataset considered for this work is summarized in Table \ref{table:virtis_obs}. 

%Surface temperature is derived from VIRTIS-M signal by modeling the 4.5-5.1 spectral radiance with a Bayesian method (Tosi et al., 2014). In general the temperature retrieved for a given point of the nucleus's surface is function of the local properties (albedo, composition, grain size, roughness, thermal conductivity, volatiles sublimation) and illumination conditions (solar incidence angle, local solar time). All points considered in this work have been acquired by VIRTIS-M around noon to early afternoon local times with spatial resolution ranging between 5.0 and 23.5 m/pixel. In these conditions the measured temperatures on the water ice-rich pixels are ranging between about 157 and 218 K.

For each spot we identified VIRTIS-M observations showing the eight spot areas imaged with the best illumination and viewing geometries to maximize the S/N ratio. The VIRTIS-M dataset considered in this work is summarized in Table 3.

Surface temperature is derived from VIRTIS-M infrared data by modeling the 4.5-5.1 $\mu m$ spectral radiance (at the pixel where the spectrum has the deepest content of H$_2$O ice) with a Bayesian approach (Tosi et al., 2014). On the surface of the nucleus, the temperature of each point is generally a function of local thermophysical properties (albedo, composition, grain size, roughness, thermal conductivity, volatiles sublimation) and instantaneous illumination conditions (solar incidence angle, or true local solar time). All measurements considered in this work were acquired by VIRTIS-M in the local solar timeframe between late morning to early afternoon, with pixel resolution ranging between 5.0 and 23.5 m/pixel. In these conditions, H$_2$O ice-rich spots show temperatures ranging between about 158 and 218 K. Since the comet's surface is not isothermal, because of   local roughness, the temperature values retrieved by VIRTIS should be considered representative only of the warmest fraction of the pixel, corresponding to the more illuminated areas. Moreover, the instrumental noise-equivalent temperature is about 150 K, corresponding to the minimum temperature detectable by the instrument. The error associated with the temperatures reported in Table 3 are between $\pm$30 K for the measurement at minimum temperature T=158 K, and $\pm$10 K for the one at maximum T=218 K. 

  %%%%%%%%
  
% \begin{figure}
  % \centering
 %\includegraphics[width=9.0cm,angle=0]{fig2.eps}
    %  \caption{ Spot N.2: as in figure 4 }
       %  \label{xx}
   %\end{figure}

\begin{table*}
\begin{center}
\caption{Parameters retrieved by modeling for each bright area. The relative error on abundance and grain size is 40\%, as estimated in Raponi 2014. }
\small{
\label{tabspots}
\begin{tabular}{c c  c  c  c c c c  } 
\hline
Spot & VIRTIS file & \multicolumn{2}{c}{Mixing modalities}  & H$_2$O ice  & H$_2$O ice grain   & Additional slope & Goodness  \\ 
& sample, line & \multicolumn{2}{c}{Alternative/Simultaneous}  & abundance(\%) & size ($\mu$m)  & (\%$\mu$m$^{-1}$) & 
$\chi^{2}$   \\ 
\hline
1 & I1$\_$00383518966         & areal    &  \multirow{2}{*}{A}  & 1.1 & 350  & \multirow{2}{*}{5.2} & 4.19  \\  
& s: 184, l: 24  & intimate &                      & 1.2 & 450  &                      & 4.16  \\ 
\hline

2 & I1$\_$00385906923         & areal    &  \multirow{2}{*}{A}  & 0.1 &  40  & \multirow{2}{*}{no}  & 1.18  \\  
& s: 173, l: 42  & intimate &                      & 0.5 & 200  &                      & 1.04  \\ 
\hline

3 & I1$\_$00385885107         & areal    &  \multirow{2}{*}{A}  & 1.3 & 750  & \multirow{2}{*}{-2.8 } & 2.78   \\  
& s: 120, l: 35  & intimate &                      & 1.8 & 1300 &                      & 3.30  \\ 
\hline

4 & I1$\_$00373462192         & areal    &  \multirow{2}{*}{S}  & 3.2 & 4500 & \multirow{2}{*}{no  } & 0.45  \\  
& s: 93, l: 35  & intimate &                      & 4.0 &   30 &                      &       \\ 
\hline

5 & I1$\_$00377182711         & areal    &  \multirow{2}{*}{A}  & 1.7 & 400  & \multirow{2}{*}{-4.0} & 1.47 \\  
& s: 61, l: 69  & intimate &                      & 2.0 & 800  &                      &  2.15 \\ 
\hline

6 & I1$\_$00376302211         & areal    &  \multirow{2}{*}{S}  & 1.9 & 6500 & \multirow{2}{*}{-1.8} & 0.28  \\  
& s: 16, l: 38  & intimate &                      & 1.0 & 250  &                       &   \\ 
\hline

7 & I1$\_$00369356914         & areal    &  \multirow{2}{*}{A}  & 1.5 &   10 & \multirow{2}{*}{-4.0} & 0.90  \\  
& s: 239, l: 66  & intimate &                      & 1.8 &  15 &                       & 0.86 \\ 
\hline

8 & I1$\_$00377184571 & areal  &  \multirow{2}{*}{A} & 0.3 & 900 & \multirow{2}{*}{no  } & 0.56 \\ 
& s: 6, l: 7   & intimate &  &   0.7  &        2200   & & 0.55 \\ 

\hline \hline
\end{tabular}
}
\end{center}
\label{tbl:virtis_parameters}
\end{table*}

\section{Spectral modeling}
   
To derive the properties of the H$_2$O ice  detected in the bright spots on the surface of the comet, a spectral analysis was performed using Hapke's radiative transfer model (Hapke, 2012), as described in Ciarniello et al. (2011). Unfortunately we are still not able to infer the real composition of the organic-rich dark terrain  present on the comet surface.  The broad absorption band centered at 3.2 $\mu$m, and the difficulty of its interpretation, have been largely discussed by Quirico et al. (2016) on the basis of  present knowledge of the composition of cometary grains and all the components available in laboratory data. 
The mixture presented in this paper was consequently modeled by means of two spectral end members: crystalline water ice, simulated by using optical constants measured at T=160 K between 1 and 4 $\mu$m (Warren et al., 1984, Mastrapa et al., 2008, Mastrapa et al., 2009, Clark et al., 2012) and a Dark Terrain unit corresponding to the average spectrum of the comet's surface after the application of photometric correction (Ciarniello et al., 2015), as shown in Fig. 3. 
The quantitative analysis is based on the spectral shape of the diagnostic absorption bands of  H$_2$O ice. The absolute level of reflectance of the model is multiplied by a free parameter to fit the data, to account for uncertainties on the radiometric and photometric accuracy as well as errors on the local geometry information, owing to unresolved shadows and roughness. 
In some cases, the measured spectra present a fictitious slope where a high signal contrast is measured between adjacent pixels, like regions near shadows. This is due to the increasing FWHM of the point spread function toward longer wavelength. To account for this effect, a slope is added to the model  to fit the measured spectrum, where it is required. Because of its artificial origin it should not be the subject of interpretation.
Before fitting, the observed spectra are corrected for spikes and instrumental artifacts.  Thermal emission is modeled and removed simultaneously to the spectral fit, as in Protopapa et al. (2014). The best-fitting result is obtained by applying the Levenberg-Marquardt method for nonlinear least squares multiple regression. During the fitting procedure, the spectral bands are weighted for the Poissonian noise as calculated in Raponi (2014). 
We have modeled areal and intimate mixing modalities: in the areal mixing, the surface is modeled as patches of pure  H$_2$O  ice and dark terrain; in the intimate mixing model the particles of the two end-member materials are in contact with each other. The model's free-parameters are the percentage and grain size of the water ice. Further details about spectral modeling are given in Filacchione et al. (2016a) and in Raponi et al. (2016).
Most of the icy regions can be described by both areal and intimate mixture, which can be alternative. In the intimate mixing case, the model always requires larger abundance and grain size than the areal mixing case to compensate for the lack of the multiple scattering contribution, which is low because of the interaction of light with the dark terrain. The two alternative solutions indicated with A in Table 4 are the extremes of a possible set of solutions in which both kind of mixtures can be present simultaneously.  For the spectra indicated with S in Table 4 (spots 4 and 6), we have only one possible solution in which the two kind of mixtures both contribute to the production of the model, and two populations of grain size are modeled at the same time, confirming the result of Filacchione et al. (2016a). Figure 4 shows spectra and images for each spot with positive identification of ice using VIRTIS-M data. The best fitting model (in red) for each spot is shown as the areal mixing case, or the areal-intimate case according to the parameters, as indicated in Table 4. 

To highlight the effect of variation in abundance and grain size of the water ice, in Fig. 5 we show,  as an example, the best fit obtained for  spot 3, in the areal mixture case, varying the H$_2$O ice abundance and the grain size. An additional slope has been set as a free parameter to obtain the best fit. The  parameters used and the resulting goodness-of-fit of the models are shown in Table 5.

 \begin{figure}
\includegraphics[width=9.0cm,angle=0]{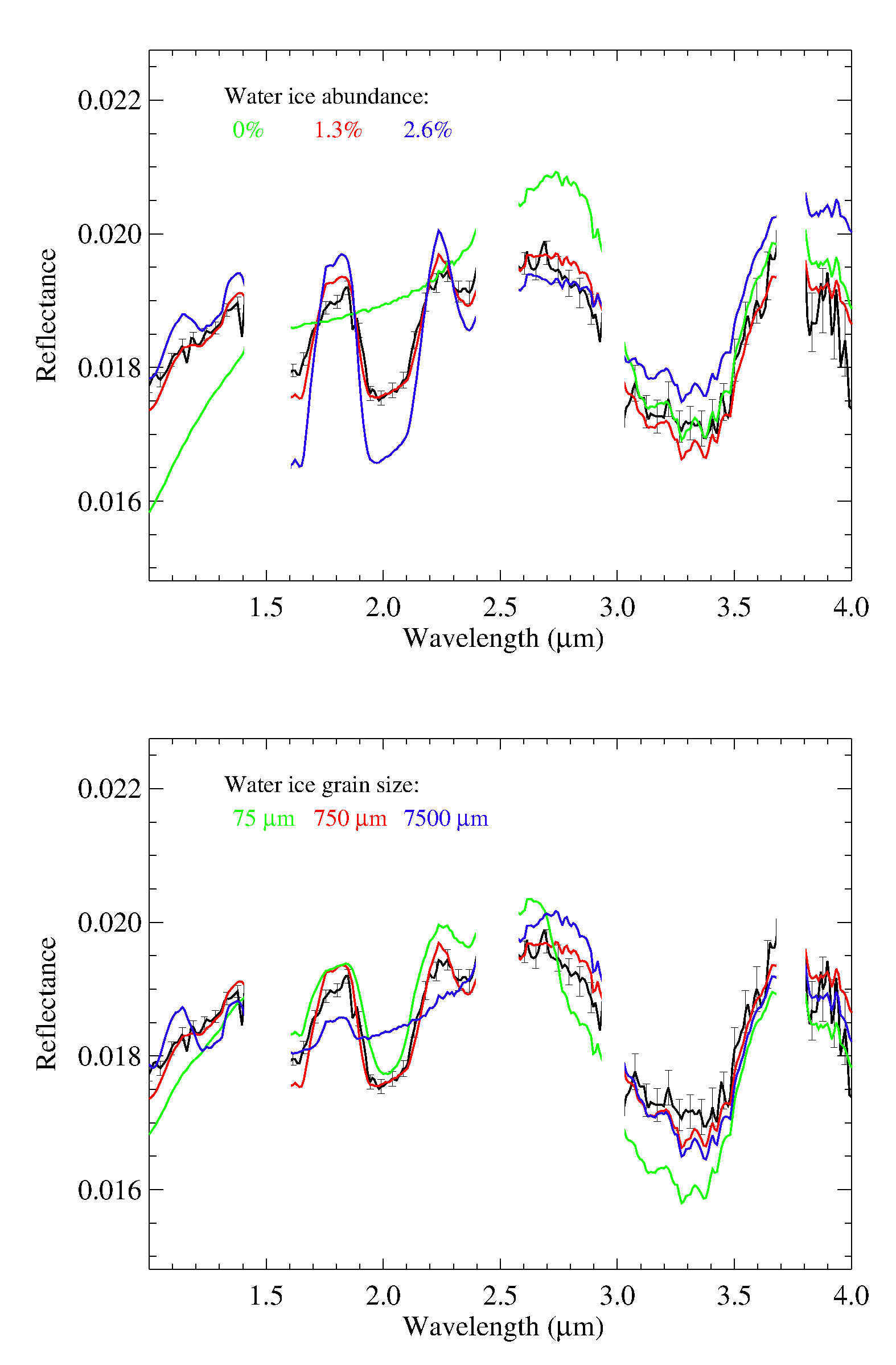}
\begin{center}
\caption {Simulated reflectance spectra are shown to highlight the effect of variation in abundance and grain size of the water ice. In both plots, the black curve is the spectrum of spot 3, and the red curve is the best fit for areal mixture case already reported in Fig. 4.  The green and blue curves are simulated by varying one of these parameters, as indicated: the panel on the top shows the effect of the variation in abundance of H$_2$O ice (0\%,  1.3\%, 2.6\%) fixing the grain size at 750$\mu$m, and the panels on the bottom show the variation in grain size (75$\mu$m, 750$\mu$m, 7500$\mu$m), fixing the H$_2$O ice abundance at 1.3\%. The gaps in the spectral ranges are not taken into account, as in Fig. 3. 
 }
 \label{xx}
 \end{center}
\end{figure}

%\begin{table}
%\begin{center}
%\caption{VIRTIS observations at the coordinates of the bright spots with the relative time of observations and number of the positive detection of ice}
%\label{virtis}

%\begin{tabular}{l l l c }
%\hline\hline
%Spot & period & observations  time & total detections  \\
%\hline
%1& STP 045 &  2015-02-24 -- 2015-03-03 & 2\\
%2&STP 048 & 2015-03-17 -- 2015-03-24 & 8\\
%&STP 049 & 2015-03-24 -- 2015-03-31 & 1\\
%3&STP 049 & 2015-03-24 -- 2015-03-31 &9\\
%&STP 051 & 2015-04-07 -- 2015-04-14 & 1\\
%&STP 053 & 2015-04-21 --  2015-04-28 & 2\\
%4&STP 026 & 2014-10-30 -- 2014-11-05 & 32\\
%5&STP 033 & 2014-12-09 -- 2014-12-16 & 86 \\
%6  & STP 013 &         2014-08-12 -- 2014-08-19  & 46\\
 %&  STP 014&   2014-08-19 -- 2014-08-26  & 45\\
  %&STP 017&    2014-09-09 -- 2014-09-16 & 25 \\
  % &STP 018&   2014-09-16 -- 2014-09-23 & 11 \\
 % &STP 026& 2014-10-30 -- 2014-11-05 &184\\
 % & STP 032&   2014-12-02 --2014-12-09 & 43\\
% & STP 045&    2015-02-24 -- 2015-03-03 &      1\\
%& STP 047      &2015-03-10 -- 2015-03-17 &     2\\
% &STP 052&     2015-04-14 -- 2015-04-21 &      7  \\
%7&STP 016 & 2014-09-02 -- 2014-09-09  & 1\\
%&STP 017 & 2014-09-09 -- 2014-09-16 & 9\\
%&STP 049 & 2015-03-24 -- 2015-03-31 & 4\\
%&STP 051 & 2015-04-07 -- 2015-04-14 & 1\\
%8 &STP 033 & 2014-12-09 -- 2014-12-16 & 3\\
%\hline

%\end{tabular}
%\end{center}
%\end{table}

 % \begin{figure}
 %\centering
%\includegraphics[width=9.0cm,angle=0]{abundance.eps}
%\caption{Spectrum of "Dark Terrain" unit corresponding to the average spectrum of the comet's surface. }
 %\label{xx}
%\end{figure}

\section{Analysis of the exposed H$_2$O ice spots }

We analyze the OSIRIS NAC images in the area where the spots have been identified and, in addition, we investigate all available observations to evaluate the lifetime of the H$_2$O ice spots. All spots are almost in equatorial or near-equatorial locations.

{\bf Spot 1.}
This bright feature is located in the southern hemisphere close to the Khonsu-Imhotep boundary and the cometary equator. The bright spot appears as a freshly exposed cliff on a rough region of rocky appearance bounding an alcove. It is likely  a remnant of a collapsed sector of a former pit. 
The first detection of this feature occurred on 5 June 2015, when it measured 57 m, whereas later on, on 27 June 2015, it measured about 36 m.

{\bf Spot 2.}
Located in the Anhur region in the southern hemisphere of the comet, this bright feature measures about 45 m. Oweing to the oblique observing geometry and many shadows cast in the neighboring terrain, it is difficult to give a clear depiction of the surroundings. Nevertheless the feature seems to correspond to a flat terrace at the centre of a roundish area (possibly a collapsed pit) in a region of consolidated materials. It shows a very low albedo for a bright feature, although it stands out from that of the typical nucleus. The OSIRIS NAC observations of this bright spot are recorded from 4 June 2015 up to 11 July 2015. As for  spot 1, this feature  is also surrounded by many shadows caused by the rugged terrain of Anhur. 

{\bf Spot 3.}
This spot appears as a bright boulder close to a pancake-like  feature (apparently composed of three broad layers), which is a morphologically unique feature right in the center of the Khonsu region (El-Maarry et al. 2016). The bright spot  has a good temporal coverage in terms of OSIRIS NAC observations. It has been observed on several occasions from the end of March to the beginning of May in 2015. The observations suggest a diminution of its size from about 18 m in late March by about 8 m in April, up to complete disappearance in May. By applying  illumination correction using the  Lommer-Seeliger disk law, the reflectance of the bright spot increased by a factor of 2 for the sequence on 25 March, unlike other cases of bright features, with the exception of spot 8. 

{\bf Spot 4.}
The OSIRIS NAC observations of 22 November 2014 reveal a cluster of bright features located  at the Atum region margin close to its boundaries with the Khonsu and Anubis regions. The bright features seem to be freshly exposed brittle materials (El-Maarry et al. 2015) in a rocky-like area. The bright patches of varying individual sizes span an area of about 25 meters in diameter. The largest patch appears to be about 10 m in size in this observation. Further analysis, including earlier images, reveals that this feature was observed as early as 2 September 2014. In the meantime the latest appearance of this feature on OSIRIS  NAC observations was on  23 November 2014. 

{\bf Spot 5.}
This bright spot faces the rounded feature with a diameter of 500 m  in Imhotep, which is interpreted by Auger et al. 2015 as a fractured accumulation basin. Similar to the bright patches of spot 4, this patch also seems to be formed by a freshly exposed area of a highly fractured material. This bright spot was observed on  22 November 2014 and measures about 10 m. It has also been  noted (Pommerol et al. 2015) that this bright feature was  spotted as early as 5 September 2014 by OSIRIS NAC. It has also been included in the Oklay et al. (2016a) study. On further analysis of the images, it is possible to find the same feature in some images on 23 August 2014, but with a smaller size.  

{\bf Spot 6.}
This is a cluster of small bright spots located in the Khepry region at the base of a scarp bordering a roundish flat terrace covered by dust deposits at the Babi region margin.  Again the material, where the bright patches are located, appears consolidated, brittle (El-Maarry et al. 2015), and dissected by pervasive fractures. The bright patches themselves seem to be either on freshly exposed outcropping material or on boulders. These spots have been observed many times from late August 2014 through to the end of November 2014 by OSIRIS NAC. The first observation on 26 August 2014 indicates the presence of around some 20 small bright spots with sizes ranging from 1.5 m to 3 m along with few spots of about 6 m in size. The following observations on 5 September reveal that there has not been any significant change in the bright patches in terms of their sizes and population. There were more observations on 16  September and three days later (Pommerol et al. 2015), suggesting the stability of the bright patches over time. Later on, an observation sequence on 29 October 2014 suggests that the cluster has  reduced to only four bright spots, each measuring about 1 m, but it is not possible to rule out the possibility that some small bright spots may be under a shadow and hence  not observable. Another sequence, on 22 November, reveals the cluster of bright spots present on observations in early September, leading to the conclusion that the observation on 29 October was subject to shadows.  This cluster of bright spots seems to have been stable for at least three months. 

Apart from this cluster of bright patches in late 2014, few observations dated from 25 March 2015 indicate the re-emergency of small bright patches at the same location. Despite the shadowy terrain, it is possible to discern up to three small bright patches, each about 5 m in diameter.  Perhaps it could be inferred that this locality near the cliff at the Khepry/Babi border is active in terms of bright patches.  Some mechanism may have triggered an outburst of icy material underneath the surface of this region allowing the cluster of bright patches to become apparent, possibly gravitational falls of boulders (Pajola et al. 2015) with consequent exposure of patches at the scarp foot.

{\bf Spot  7.}
This is a cluster of small bright patches that might correspond to a series of boulders at the base of a small terrace bounded by scarps. This feature appears  on the OSIRIS NAC observations of 30 September 2014. It is located on a slope adjacent to the smooth plain in Imhotep, pointing to the neighboring Apis region. The location itself is somewhat shadowy and is camouflaged by the surrounding terrain, making it challenging to observe the full extension of this feature. VIRTIS-M data support the presence of H$_2$O ice for this bright feature located in Imhotep region with a size of about 5 m. The corresponding positive VIRTIS-M observations date back to early September 2014, suggesting that this bright feature has been on the cometary surface even as early as the beginning of September. Therefore this feature could have a lifespan of at least one month. 

{\bf Spot 8.}
This Imhotep-based bright feature is recorded in several epochs.  The bright spot is close to the isolated accumulation basin and to the small roundish features interpreted by Auger et al. 2015 as ancient degassing conduits.  It is located at the base of what appears to be an open trench surrounded by steep small scarps. Therefore the full view of this feature is somewhat hampered by the constant casting of shadows and the viewing geometry. Nevertheless the multiple observations offer partial views of the feature that seems to lie on a consolidated flat area with fractures and tiny staircase borders. For example on the image recorded on 30 September 2014, it appears to  be composed of two segments with one measuring 3 m, while the other measures 6 m approximately. We note that the absolute reflectance of this bright spot increased by a factor of 3 upon applying the illumination correction using Lommer-Seeliger disk law for the observation sequence on 5 September 2014. A similar effect has only been noticed in the case of  spot 3, where the factor was 2.   The earliest detection of this feature dates to 25 August 2014 and the corresponding positive VIRTIS-M recordings date to mid-December 2014, suggesting a stable existence of almost four months on surface for this bright feature.

\begin{table}
\begin{center}
\caption{Parameters used to perform the models shown in Fig. 5. The first line represents the best fit (reported in Table 4) shown in red in both panels of Fig. 5.    The abundance and grain size of water ice in areal mixture are fixed as: 0\% and twice the abundance retrieved for the best fit, and one tenth and ten times the grain size retrieved for the best fit. The additional slope (free parameter of the model) and the resulting goodness are indicated in the table.}
\small{
\label{tabspots}
\begin{tabular}{c c  c  c  }
\hline
H$_2$O ice  & H$_2$O ice grain   & Additional slope & Goodness  \\ 
 abundance(\%) & size ($\mu$m)  & (\%$\mu$m$^{-1}$) & $\chi^{2}$   \\ 
\hline
\hline
1.3 & 750 & -2.8 & 2.78  \\  
\hline
0 & 750 & -4 &  31.2  \\  
\hline
2.6  & 750 &  0 &  15.0  \\ 
\hline
1.3  & 75 &  -2.5 &  9.4  \\ 
\hline
1.3  & 7500 &  -3.9 &  6.8 \\ 
\hline
\hline 
\end{tabular}
}
\end{center}
\label{tbl:virtis_parameters}
\end{table}

\section{Temporal evolution of the bright spots}

Although only a subset of the bright spots detected by OSIRIS can be analyzed by VIRTIS, the unambiguous detection of the spectral signatures of H$_2$O ice in eight of these bright spots is a clear confirmation of their icy nature. In section 5, we detail evidence of spots with long life time. Here (Fig. 6)  we add a comparison for a cluster of bright spots (spot 6) in Khepry  with observations at an interval longer than two months, which confirms the stability of this cluster at that time of the mission.  

From the VIRTIS and OSIRIS data, we can measure several parameters: estimation of the amount of water ice on each spot, its local temperature, a timescale and an extent of erosion. These were measured at different times and should consequently be considered only as first order indications. They are theoretically linked together, so that we can use them to estimate whether ice behaves as expected for each spot. In Table 6 we report the mass release rate of H$_2$O (Prialnik et al. 2004), for each temperature, from the surface of 67P/C-G estimated in each of these locations as follows

\begin{equation}
\label{mass_release}
%\mathcal{Q}_{H_2O} = \frac{{P}_{H_2O}(T) } {\sqrt{2 \pi m_{H_2O} {k_B T}} } 
\mathcal{Q}_{H_2O} = \mathcal{P}_{H_2O} (T) \sqrt{\frac{m_{H_2O}}{2 \pi k_B T}}
,\end{equation}
with $m_{H_2O}$ [kg] the mass of one molecule of H$_2$O, k$_B$ the Boltzmann constant, T [K] the temperature of each spot, and $\mathcal{P}_{H_2O}$ the saturation vapor pressure, which can be written as 

\begin{equation}
\label{P}
\mathcal{P}_{H_2O} = A e^{-B/T}
,\end{equation}

with A=356~10$^{10}$~Nm$^{-2}$ and B=6141.667~K for water (Fanale and Salvail, 1984).

\begin{table}
\begin{tiny}
\caption{\label{QH2O}Mass release rate of H$_2$O from the surface of 67P/C-G at the location of H$_2$O ice-rich spots. In the last column an indication of the observed lifetime of the spots is summarised.}
\begin{tabular}{cccl}
\hline\hline
Spot & T [K]     &   $\mathcal{Q}_{H_2O}$ [kg m$^{-2}$s$^{-1}$] & OSIRIS observations \\

\hline
1 & 203 & 3.357 $\times$ 10$^{-4}$  &  size receded from 57m \\
 & & &  to 36m in 3 weeks\\
2 & 197 & 1.336 $\times$ 10$^{-4}$  & observed for 5 weeks\\
3 & 218 & 2.485 $\times$ 10$^{-3}$ & size receded from 18m \\
& & & to 8m in 3 weeks,  disappear \\
 & & &    in the following 3 weeks \\
4 & 168 & 0.662 $\times$ 10$^{-6}$  & cluster stable for 11 weeks\\
5 & 188 & 3.003 $\times$ 10$^{-5}$ & stable for 13 weeks\\
6 & 179 & 0.625 $\times$ 10$^{-5}$ & cluster stable for 13 weeks\\
7 & 163 & 2.156 $\times$ 10$^{-7}$ & cluster stable for 3 weeks\\
8 & 158 & 0.701 $\times$ 10$^{-7}$ & stable for 15 weeks\\
\hline

\end{tabular}\\
\end{tiny}
\end{table}

~

For each spot, the temperature changes because of local diurnal variations of the solar input, seasonal effects, shadowing, or self-heating. These effects are difficult to estimate since they depend on the local illumination geometry and thermo-physical properties. A low thermal inertia, as derived by VIRTIS (Capaccioni et al. 2015) and MIRO (Schloerb et al. 2015), results in quite large day-night and seasonal variations of the temperature.  These may influence the survival of ice-rich spots in a way that may be unpredictable. However, we find that the behavior of the eight spots found by the OSIRIS-VIRTIS study is in good agreement with the expected thermal behavior of H$_2$O ice. 
\par
Although the size and timescales measured for each spot may be considered with caution owing to the errors like those induced by shadows,  they allow for another estimate of the mass release rate (Prialnik et al. 2004) through $\mathcal{Q}_{H_2O} = \varrho ~ \Delta l / \Delta t$, with $\varrho$ the local density of water ice, $\Delta l$ the typical extent of erosion of each spot, and $\Delta t$ its erosion timescale. If we take the example of spot 1, the expected timescale to erode this feature at the observed extent and for a temperature of 203~K would be 50-100~hr. We emphasize  that given the numerous uncertainties this should be considered as a first order approximation. However, this timescale is such that the feature should be stable against day-night variations of the temperature, since it would take more than one cycle to erode it. In addition, the lower temperatures of the night would prevent such a rapid erosion.
The total mass of water released by the erosion of spot 1 is $M=\mathcal{Q}_{H_2O} \Delta t f \Delta S$ with $\Delta S$ the eroded surface and $f$ the water ice fraction inferred from VIRTIS data modeling. 
The feature containing 1\% of water ice seems to have decreased from 57~m to 36~m in about 22~days, i.e. $\sim$1500m$^2$ were eroded in $\sim$2$\times$10$^6$s.  At 203~K, this translates into $\sim$10$^4$kg of water ice being sublimated, which is not possible to reach with surface ice alone. We thus have to assume  that 1) icy grains may be scattered to the nucleus surface and recondence, creating a cycle maintaining water ice at the surface (Crifo, 1987; Davidsson and Skorov, 2004), 2) the average temperature is much lower than the temperature measured by VIRTIS, 3) ice-rich subsurface layers  contribute to maintaining the surface ice. Case 1 is beyond the scope of the simple calculations we perform here. For case 2, we estimate that surface ice (contained in a 1mm layer) would be able to reproduce the observed behavior if the average temperature is $\sim$175~K: large day-night variations of the temperature would thus be necessary. For case 3, if we assume that subsurface layers have the same properties as the bulk of the comet, a layer of the order of $\sim$10 cm would be required to explain the observed mass release.  We note that in 67P/C-G, the diurnal and seasonal skin depths ($ \left( \frac{P_{spin} \kappa}{\pi \rho c}  \right)^{1/2}  $ and $ \left( \frac{2 \kappa a^{3/2}}{\sqrt{GM_{\odot}} \rho c}  \right)^{1/2}  $, with a the semi major axis, c the specific heat, G the gravitational constant, $\kappa$ the thermal conductivity, $M_{\odot}$ the solar mass, $P_{spin}$ the spin period, and $\rho$ the bulk density; Prialnik et al. 2004) vary from 1~mm to 9~cm and 80~cm to 7~m respectively, depending on local thermo-physical properties, as computed by Leyrat et al. (2015). 
 
Spot 2 is larger than spot 1, but contains an order of magnitude less H$_2$O ice as inferred from spectral modeling. This feature's evolution is found to be in good agreement with the expected thermal behavior of water ice, as well as the inference from its low albedo that this spot is close to the end of the sublimation phase.
 
Spot 3 is the hottest feature measured in this study. If we assume that the ice-rich boulder has the same properties as the bulk of 67P/C-G, the 18~m feature being eroded in $\sim$3$\times$10$^6$s results in a mass release rate of 2.82$\times$10$^{-5}$kg~m$^{-2}$~s$^{-1}$, i.e. two orders of magnitude less than the rate computed from the measured temperature. We should thus assume that the temperature is much lower most of the time at this spot, or the boulder properties vary from those of the bulk of the comet (a local density of 800 kg~m$^{-3}$ for example would be required). Alternatively, the very high temperature observed for spot 3 is likely the result of an areal mixture of icy and ice-free terrain within the VIRTIS pixel. Indeed, for a given observed albedo and temperature, the mixing of ice and dust at a grain level plays an important role: ice intimately mixed with dust will be hotter and shorter-lived than a patch of pure ice surrounded by dust.
Given the low temperatures encountered for spots 4 to 8, it is expected that these features should be long-lived, as  observed by OSIRIS and VIRTIS.   At a lower temperature of 180K (spot 6), the sublimation rate is only 0.625$\times$10$^{-5}$kg~m$^{-2}$~s$^{-1}$, and decreases exponentially at lower temperatures (spots 4, 7, and 8).

The appearance of meter-sized spots, which are mostly only illuminated for a short fraction of the day, remains constant over time.  These features would be more affected by seasonal variations than diurnal variations of the temperature, since water ice is mostly stable at the measured temperatures. Cometary activity, triggered below the surface by other volatile species, may locally influence the surface properties, such as the distribution of dust, and expose fresh H$_2$O ice at the surface. 
 
All the bright spots are on consolidated dust free materials, either on boulders or on freshly exposed outcropping regions that often display penetrative fractures. This suggests that H$_2$O ice can mainly be  found on the consolidated  substratum exposed along scarps or detached in the form of boulders.  Some of the bright spots  have been in place for weeks and months, while others seem related to diurnal variation.

\begin{figure}
 \centering
\includegraphics[width=9.0cm,angle=0]{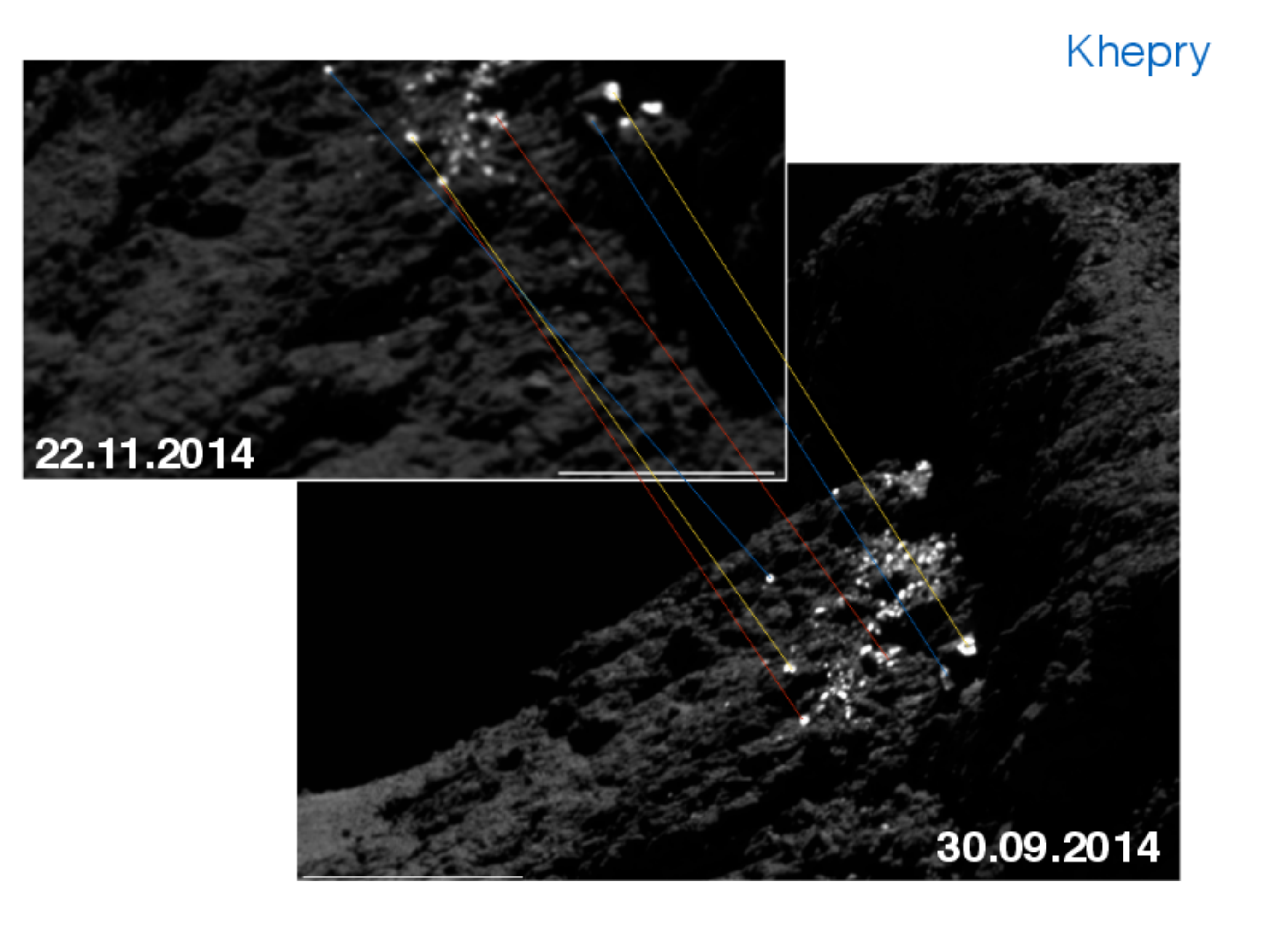}
\caption{Comparisons of OSIRIS NAC images of the cluster of bright spot 6,  observed two months apart, which shows the stability of the bright features with time. }
 \label{xx}
\end{figure}

\section{Conclusions}

Comet 67P/Churyumov-Gerasimenko  shows a surface rich in heterogeneous geological structures and surface morphological variations that show color and albedo variations across the surface. The high-resolution images obtained by OSIRIS enable us to identify a large quantity of bright spots of different size and located in areas with different properties and high albedo.  

In this paper, we present for the first time a complementary study of data acquired by the OSIRIS and VIRTIS instruments. A major objective of this paper is to firmly detect the presence of H$_2$O ice on the comet's surface. We confirm the presence of H$_2$O ice on eight new spots and we model the spectra with   H$_2$O ice and dark material.  

Comparing the coordinates of the detected eight H$_2$O ice spots with those of 67P/C-G dust jets,  five spots  (4, 5, 6, 7 and 8) have been found to lie in the same approximate position of the jets identified by Vincent et al. (2016a) and one (spot 3) among the outbursts observed in the cometary summer (Vincent et al., 2016b). Observational evidence showed that the majority of dust jets also arose from rough terrains and fractured walls rather than smooth areas (Vincent et al., 2016a). Some of these detected H$_2$O ice spots have also been compared by Oklay et al. (2016b) to those of comets 9P and 103P.

The detection of  H$_2$O ice signatures by VIRTIS on eight of the 13  locations  given by OSIRIS data does not mean that the other spots do not contain  ice on their surface and this can be explained by not simultaneous observations, unfavorable instrumental signal-to-noise conditions, spatial resolution on the surface, different illumination/viewing geometry, and by the fact that VIRTIS-M channel was unavailable  after 4 May, 2015 owing to the failure of the active cooler. 

~

The main results of this work can be summarized as follows:
\begin{itemize}

\item We presented for the first time a complementary analysis of H$_2$O ice-rich areas using data acquired by the OSIRIS and VIRTIS instruments. Comparing high spatial resolution VIS images with extended IR range spectra enables us  to study  the morphological, thermal and compositional properties of these areas at the same time. 
\item The analysis of the spectral properties observed by VIRTIS-M indicates that, on these areas, the H$_2$O ice abundance is between 0.1 and 7.2$\%$, mixed in areal or/and in intimate modalities with the dark terrain. 
\item The ice is distributed on the two lobes of 67P/C-G in locations which remain  in shadow for longer.
\item The detected bright spots are mostly on consolidated dust free material surfaces,  mostly concentrated in equatorial latitudes.
\item {The mass release of H$_2$O at the location of the eight ice-rich spots has been estimated.}
\item Some spots are stable for several months and others show temporal changes connected to diurnal and seasonal variations. Stability of the spots is corroborated by the temperature retrieved at the surface. The behavior of ice on these locations is in very good agreement with theoretical expectations.
\item Six of the detected H$_2$O ice spots are located in  approximately the same position of the previously detected cometary jets.
%\item there is a strong correlation between the occurrence of the 2-\B5m (evidence of H2O deposits) band and a flat VIS slope enabling future joint Osiris-Virtis studies for this purpose 
\end{itemize}

H$_2$O ice is present on the surface substratum where solar illumination plays an important role with seasonal and diurnal variations. During the perihelion orbit passage of the comet, the Rosetta spacecraft was at a greater distance and the available surface OSIRIS images were at lower resolution.  Starting in March 2016, the comet is observed again from close distances.  With  analysis of other available data (in particular from OSIRIS), we will study the surface changes after the perihelion passage to better understand  the surface evolution of the comet.\\

\begin{acknowledgements}
OSIRIS was built by a consortium of  the Max-Planck-Institut f\"ur
Sonnensystemforschung, G\"ottingen, Germany, CISAS--University of
Padova, Italy, the Laboratoire d'Astrophysique de Marseille, France, the
Instituto de Astrof\'isica de Andalucia, CSIC, Granada, Spain, the Research and
Scientific Support Department of the European Space Agency, Noordwijk, The
Netherlands, the Instituto Nacional de T\'ecnica Aeroespacial, Madrid, Spain,
the Universidad Polit\'echnica de Madrid, Spain, the Department of Physics and
Astronomy of Uppsala University, Sweden, and the Institut  f\"ur Datentechnik
und Kommunikationsnetze der Technischen Universit\"at  Braunschweig, Germany.

VIRTIS was built by a consortium from Italy, France, and Germany, under the scientific responsibility of IAPS, Istituto di Astrofisica e Planetologia Spaziali of INAF, Rome, which lead also the scientific operations.  The VIRTIS instrument development for ESA has been funded and managed by ASI (Italy), with contributions from Observatoire de Meudon (France) financed by CNES and from DLR (Germany). The VIRTIS instrument industrial prime contractor was former Officine Galileo, now Finmeccanica in Campi Bisenzio, Florence, Italy. 

The support of the national funding agencies of Germany (DLR), France (CNES),
Italy (ASI), Spain (MEC), Sweden (SNSB), and the ESA Technical Directorate is
gratefully acknowledged. 

\end{acknowledgements}

\end{document}